\documentclass[12pt,a4]{article}

\usepackage{graphicx,amsmath,amsthm,bm,fancybox,ascmac,amssymb,multirow,hhline,atbegshi}

\usepackage{natbib}
\usepackage{verbatim}
\usepackage{color}

\usepackage[dvipdfmx]{hyperref}
\hypersetup{
 setpagesize=false,
 bookmarksnumbered=true,%
 bookmarksopen=true,%
 colorlinks=true,%
 linkcolor=red,
 citecolor=blue,
}

\makeatletter
\def\section{\@startsection {section}{1}{\z@}{-3.5ex plus -1ex minus -.2ex}{2.3 ex plus .2ex}{\large\sc\centering}}
\def\subsection{\@startsection {subsection}{1}{\z@}{-3.5ex plus -1ex minus -.2ex}{2.3 ex plus .2ex}{\large}}
\makeatother

\theoremstyle{definition}

\newtheorem{thm}{Theorem}
\newtheorem{lem}{Lemma}

\title{\large\bf On the Consistency of the Bias Correction Term of the AIC for the Non-Concave Penalized Likelihood Method}
\author{\normalsize Yuta Umezu\thanks{Corresponding author.
Graduate School of Mathematics, Kyushu University. 744 Motooka, Nishi-ku, Fukuoka 819-0395, Japan.
Email: y-umezu@math.kyushu-u.ac.jp}
\and \normalsize Yoshiyuki Ninomiya\thanks{
Institute of Mathematics for Industry, Kyushu University. 744 Motooka, Nishi-ku, Fukuoka 819-0395, Japan.
}
}
\date{\normalsize Version: \today}

\selectfont

\allowdisplaybreaks

\begin{document}
\maketitle
\begin{abstract}
Penalized likelihood methods with an $\ell_{\gamma}$-type penalty, such as the Bridge, the SCAD, and the MCP, allow us to estimate a parameter and to do variable selection, simultaneously, if $\gamma\in (0,1]$.
In this method, it is important to choose a tuning parameter which controls the penalty level, since we can select the model as we want when we choose it arbitrarily.
Nowadays, several information criteria have been developed to choose the tuning parameter without such an arbitrariness.
However the bias correction term of such information criteria depend on the true parameter value in general, then we usually plug-in a consistent estimator of it to compute the information criteria from the data.
In this paper, we derive a consistent estimator of the bias correction term of the AIC for the non-concave penalized likelihood method and propose a simple AIC-type information criterion for such models.

\

\end{abstract}

\section{Introduction}
\label{sec_intro}
Sparse regularization method, such as the Lasso (\citealt{Tib96}), the SCAD (\citealt{FanLi01}), the MCP (\citealt{Zha10}), and the Bridge (\citealt{FraFri93}), is a kind of estimation method by imposing a penalty term which is not differentiable at the origin on an estimating function, and these methods allow us to estimate a parameter and to do variable selection, simultaneously.
The estimator obtained from such methods, in general, depends on a so-called tuning parameter $\lambda$ which is a positive constant controling a penalty level.
Concretely speaking, if $\hat{\bm{\beta}}_{\lambda}=(\hat{\beta}_{\lambda,1},\hat{\beta}_{\lambda,2},\ldots,\hat{\beta}_{\lambda,p})^{{\rm T}}$ is the estimator based on such a method, several of its components will be shrunk to exactly 0 when $\lambda$ is not close to 0.
While the Lasso can be solved by using convex optimization problem if an estimating function is convex and is useful, the efficiency of the parameter estimation based on the Lasso is not necessarily large, because the Lasso shrinks the estimator to the zero vector too strongly.
To avoid such a problem, it has been proposed to use a penalty term that does not shrink the estimator with the large value.
Typical examples of such regularization methods are the Bridge, the SCAD, and the MCP.
Whereas the Bridge uses an $\ell_{\gamma}$ penalty term ($0<\gamma<1$), the SCAD and the MCP use penalty terms which can be approximated by an $\ell_{1}$ penalty term in the neighborhood of the origin and which we hereafter call an $\ell_{1}$-type.
Although it is difficult to obtain their estimates since their penalties are non-convex, there are several algorithms which assure the convergence to a local optimal solution such as the coordinate descent method and the gradient descent method.

On the other hand, we have to choose the proper value of $\lambda$ which exists in the above regularization methods, and this is an important task for the appropriate model selection.
One of the simplest methods for selecting $\lambda$ is to use the cross-validation (CV; \citealt{Sto74}).
A stability selection method (\citealt{MeiBuh10}) based on subsampling in order to avoid problems caused by selecting a model based on only one value of $\lambda$ is attractive, but it requires a considerable number of computational costs like the CV.
For an analytical approach without such a problem, information criteria have been developed rapidly (\citealt{YuaLin07,WanLiTsa07,WanLiLen09,ZhaLiTsa10,FanTan13}).
Letting $\ell(\cdot)$ be the log-likelihood function and $\hat{\bm{\beta}}_{\lambda}$ be the estimator of $\bm{\beta}$ obtained by the above regularization methods, their information criteria take the form of $-2\ell(\hat{\bm{\beta}}_{\lambda})+\kappa_{n}\|\hat{\bm{\beta}}_{\lambda}\|_{0}$.
Then, for some sequence $\kappa_{n}$ that depends on at least the sample size $n$, the model selection consistency is at least assured.
For example, the information criterion with $\kappa_{n}=\log n$ is proposed as the BIC.
This approach includes the results for the case in which the dimension of the parameter vector $p$ goes to infinity, and so it has a high value.
However, the choice of the tuning parameter remains somewhat arbitrary.
That is, there is a class of $\kappa_{n}$ assuring a preferred asymptotic property such as the model selection consistency, but there are no appropriate rules to choose the one from the class.
For example, since the BIC described above is not derived from Bayes factor, there is no reason to use $\kappa_n=\log n$ instead of $\kappa_n=2\log n$.
This is a severe problem because data analysts can choose $\kappa_{n}$ arbitrarily and do model selection as they want.

The information criterion without such a problem about the arbitrariness is proposed by \citet{EfrHasJohTib04} or \citet{ZouHasTib07} in a Gaussian linear regression setting or by \citet{NinKaw14} in a generalized linear regression setting.
Concretely speaking, based on the original definition of the $C_{p}$ or the AIC, an unbiased estimator of the mean squared error or an asymptotically unbiased estimator of a Kullback-Leibler divergence is derived.
However, these information criteria are basically only for the Lasso.
In addition, the asymptotic setting used in \citet{NinKaw14} does not assure even the estimation consistency.
\cite{UmeNin15} generalize the above result to non-concave penalized likelihood by the same argument as in \cite{NinKaw14} and derive an asymptotically unbiased estimator of a Kullback-Leibler divergence, i.e., bias correction term, although they do not consider the consistency of the bias correction term.
In fact, it is difficult to construct a consistent estimator of the bias correction term in their setting especially for an $\ell_{1}$-type regularization methods such as SCAD and the MCP.

To derive an unbiased estimator of a bias correction term in the AIC, we focus on a so-called oracle property.
Roughly speaking, an oracle property consists of a sparsity of the estimator corresponding to the zero part of the true parameter and an asymptotic normality of the estimator corresponding to the non-zero part of the true parameter.
Instead of the sparsity, a variable selection consistency of the estimator corresponding to the non-zero part of the true parameter is often refered in the oracle property (see, e.g., \citealt{FanLi01,Zou06}). 
The definition of an oracle property is given in Section \ref{sec_Asymptotics}.

Our goal in this paper is to consider an asymptotic property of the estimator obtained from the regularization method especially in the case of $\ell_{1}$-type penalty and the Bridge for $\ell_{\gamma}$-type penalty.
Moreover we will derive an information criterion based on such estimators by the same argument as in \cite{UmeNin15}.
To do this, we assume that the tuning parameter converges to 0 with proper rate as $n$ goes to infinity.
Unfortunately, to assure the above mentioned properties, the rate of decay of $\lambda\;(=\lambda_{n})$ is deferent in $0<\gamma<1$ and $\gamma=1$.
In fact, $n^{-1/2}\lambda_{n}\to 0$ and $n^{-1/2}\lambda_{n}\to\infty$ are required for $\ell_{\gamma}$-type penalty $(0<\gamma<1)$ and $\ell_{1}$-type penalty, respectively.
Therefore, we need to divide into two cases, i.e., whether $n^{-1/2}\lambda_{n}$ converges or not.
Then, for the Kullback-Leibler divergence, we construct an asymptotically unbiased estimator by evaluating the asymptotic bias between the divergence and the log-likelihood in which the estimator is plugged.

The rest of the chapter is as follows.
In Section \ref{sec_Model}, the generalized linear model and the $\ell_{1}$-type regularization method we treat are introduced, and then some assumptions for our asymptotic theory are described.
In Section \ref{sec_Asymptotics}, we discuss an asymptotic property of the estimator obtained from the regularization method, and in Section \ref{sec_ic}, we use the asymptotic property to evaluate an asymptotic bias, which is needed to derive the AIC.
The result for the Bridge estimator is mentioned in Section \ref{sec_aic_bridge}.
Some concluding remarks and future works are presented in Section \ref{sec_conclusion} and several proofs are relegated to Appendix.

\section{Setting and assumptions for asymptotics}\label{sec_Model}
Let us consider a natural exponential family with a natural parameter $\bm{\theta}$ in $\Theta\;(\subset\mathbb{R}^{r})$ for an $r$-dimensional random variable $\bm{y}$, whose density is 
\begin{align*}
f(\bm{y};\bm{\theta})=\exp\left\{\bm{y}^{{\rm T}}\bm{\theta}-a(\bm{\theta})+b(\bm{y})\right\}
\end{align*}
with respect to a $\sigma$-finite measure. We assume that $\Theta$ is the natural parameter space; that is, $\bm{\theta}$ in $\Theta$ satisfies $0<\int\exp\{\bm{y}^{{\rm T}}\bm{\theta}+b(\bm{y})\}d\bm{y}<\infty$. Accordingly, all the derivatives of $a(\bm{\theta})$ and all the moments of $\bm{y}$ exist in the interior $\Theta^{\circ}$ of $\Theta$, and, in particular, ${\rm E}[\bm{y}]=a'(\bm{\theta})$ and ${\rm V}[\bm{y}]=a''(\bm{\theta})$. For a function $c(\bm{\eta})$, we denote $\partial c(\bm{\eta})/\partial \bm{\eta}$ and $\partial^{2} c(\bm{\eta})/\partial \bm{\eta}\partial \bm{\eta}^{{\rm T}}$ by $c'(\bm{\eta})$ and $c''(\bm{\eta})$, respectively. We also assume that ${\rm V}[\bm{y}]=a''(\bm{\theta})$ is positive definite, and hence, $-\log f(\bm{y};\bm{\theta})$ is a strictly convex function with respect to $\bm{\theta}$.

Let $(\bm{y}_{i},\bm{X}_{i})$ be the $i$-th set of responses and regressors $(i=1,2,\ldots,n)$; we assume that $\bm{y}_{i}$ are independent $r$-dimensional random vectors and $\bm{X}_{i}$ in ${\cal X}\;(\subset\mathbb{R}^{r\times p})$ are $(r\times p)$-matrices of known constants. We will consider generalized linear models with natural link functions for such data (see \citealt{McCNel83}); that is, we will consider a class of density functions $\{f(\bm{y};\bm{X}\bm{\beta});\;\bm{\beta}\in{\cal B}\}$ for $\bm{y}_{i}$; thus, the log-likelihood function of $\bm{y}_{i}$ is given by
\begin{align*}
g_{i}(\bm{\beta})=\bm{y}_{i}^{{\rm T}}\bm{X}_{i}\bm{\beta}-a(\bm{X}_{i}\bm{\beta})+b(\bm{y}_{i}),
\end{align*}
where $\bm{\beta}$ is a $p$-dimensional coefficient vector and ${\cal B}\;(\subset\mathbb{R}^{p})$ is an open convex set. To develop an asymptotic theory for this model, we assume two conditions about the behavior of $\{\bm{X}_{i}\}$, as follows:
\begin{itemize}
\item[(C1)] ${\cal X}$ is a compact set with $\bm{X}\bm{\beta}\in\Theta^{\circ}$ for all $\bm{X}\ (\in{\cal X})$ and $\bm{\beta}\ (\in{\cal B})$.
\item[(C2)] There exists an invariant distribution $\mu$ on ${\cal X}$. In particular, $n^{-1}\sum_{i=1}^{n}\bm{X}_{i}^{{\rm T}}a''(\bm{X}_{i}\bm{\beta})\bm{X}_{i}$ converges to a positive-definite matrix $\bm{J}(\bm{\beta})\equiv\int_{{\cal X}}\bm{X}^{{\rm T}}a''(\bm{X}\bm{\beta})\bm{X}\mu({\rm d}\bm{X})$.
\end{itemize}
In the above setting, we can prove the following lemma.
\begin{lem}
Let $\bm{\beta}^{*}$ be the true value of $\bm{\beta}$. Then, under conditions (C1) and (C2), we obtain the following:
\begin{itemize}
\item[(R1)] There exists a convex and differentiable function $h(\bm{\beta})$ such that $n^{-1}\sum_{i=1}^{n}\{g_{i}(\bm{\beta}^{*})-g_{i}(\bm{\beta})\} \stackrel{{\rm p}}{\to} h(\bm{\beta})$ for each $\bm{\beta}$.
\item[(R2)] $\bm{J}_{n}(\bm{\beta})\equiv -n^{-1}\sum_{i=1}^{n}g''_{i}(\bm{\beta})$ converges to $\bm{J}(\bm{\beta})$.
\item[(R3)] $\bm{s}_{n}\equiv n^{-1/2}\sum_{i=1}^{n}g'_{i}(\bm{\beta}^{*})\stackrel{{\rm d}}{\to}\bm{s}\sim {\rm N}(\bm{0},\bm{J}(\bm{\beta}^{*}))$.
\end{itemize}
\end{lem}
\noindent
See \citet{NinKaw14} or \citet{UmeNin15} for the proof.

First, we consider an $\ell_{1}$-type regularized estimator such as the SCAD and the MCP.
Let us consider a non-concave penalized maximum likelihood estimator
\begin{align}
\hat{\bm{\beta}}_{\lambda}=
\underset{\bm{\beta}\in{\cal B}}{{\rm argmin}}\left\{-\sum_{i=1}^{n}g_{i}(\bm{\beta})+n\sum_{j=1}^{p}\eta_{\lambda_{n}}(\beta_{j})\right\},
\label{eq;est_n}
\end{align}
where for $\lambda\;(>0)$, $\lambda_{n}=n^{(\gamma_{0}-2)/2}\lambda$ is a tuning parameter depending on $n$ and $\eta_{\lambda_{n}}(\beta_{j})$ is a penalty term with respect to $\beta_{j}$, which is not necessarily convex.
In addition, we assume that $\gamma_{0}\in[1,2)$.
Note that $n^{1/2}\lambda_{n}=\lambda$ and $n^{1/2}\lambda_{n}\to\infty$ when $\gamma_{0}=1$ and $\gamma_{0}\in(1,2)$, respectively, and $\lambda_{n}\to 0$ when $\gamma_{0}\in[1,2)$.

To develop the asymptotic property of the estimator obtained from (\ref{eq;est_n}), we need to set several conditions for $\ell_{1}$-type penalty:
\begin{itemize}
\item[(P1)]
$\eta_{\lambda_{n}}(\beta)$ is not differentiable only at the origin, symmetry with respect to $\beta=0$, and monotone non-decreasing with respect to $|\beta|$.
\item[(P2)]
$\lim_{\beta\to0}\eta_{\lambda_{n}}(\beta)/|\beta|=\lambda_{n}$.
\item[(P3)]
$\lim_{n\to\infty}\eta_{\lambda_{n}}(\beta)=0$ for any $\beta$.
\item[(P4)]
There exist $\tau\;(>0)$ such that $\eta'_{\lambda_{n}}(\beta)=0$ for any $|\beta|\geq \tau \lambda_{n}$.
\item[(P5)]
$\lim_{n\to\infty}\eta''_{\lambda_{n}}(\beta)=0$ for any $\beta\;(\neq 0)$.
\end{itemize}
The conditions (P1) and (P2) are almost the same conditions as in \citet{UmeNin15}, and (P2) implies $\eta_{\lambda_{n}}(0)=0$ and
\begin{align}
\lim_{\beta\to0}\eta'_{\lambda_{n}}(\beta)/{\rm sgn}(\beta)=\lambda_{n}
\label{eq;dif}
\end{align}
(P3) is needed to assure the consistency of the estimator.
The condition (P4) guarantees the asymptotic unbiasedness of the estimator in (\ref{eq;est_n}).
Note that these conditions are still hold for the SCAD, and the MCP although the Lasso does not satisfy the condition (P4) since ${\rm d}|\beta|/{\rm d}\beta={\rm sgn}(\beta)$ for any $\beta\;(\neq 0)$.

\citet{NinKaw14} consider the same model although their model is considered only for the Lasso with $\gamma_{0}=2$.
However, even the consistency of the estimator does not assured in their setting because the penalty term does not vanish as $n\to\infty$.
On the other hand, \citet{UmeNin15} put $n^{1/2}$ on the penalty term for the non-concave penalty with $\gamma_{0}=2$ to avoid such a problem although the asymptotic distribution of the estimator has an asymptotic bias.
We put $n$ on the penalty term and consider the case of $\gamma_{0}\in[1,2)$.
From this, we can prove the oracle property of the estimator when $\gamma_{0}\in(1,2)$ although the same result as in \citet{UmeNin15} still hold when $\gamma_{0}=1$.

\section{Asymptotic behavior}
\label{sec_Asymptotics}

\subsection{Preparation}
Because the penalty term in (\ref{eq;est_n}) converges to 0, we can immediately see that the following lemma holds by the same argument of \citet{KniFu00} or \citet{UmeNin15}:
\begin{lem}[consistency]\label{lem;consistency}
$\hat{\bm{\beta}}_{\lambda}$ is a consistent estimator of $\bm{\beta}^{*}$ under the conditions (C1),\;(C2), and (P1)--(P3).
\end{lem}

Hereafter, we will denote $\bm{J}(\bm{\beta}^{*})$ by $\bm{J}$ so long as there is no confusion. In addition, we denote $\{j;\;\beta^{*}_{j}=0\}$ and $\{j;\;\beta^{*}_{j}\neq 0\}$ by ${\cal J}^{(1)}$ and ${\cal J}^{(2)}$, respectively. Moreover, the vector $(u_{j})_{j\in{\cal J}^{(k)}}$ and the matrix $(\bm{J}_{ij})_{i\in{\cal J}^{(k)},j\in{\cal J}^{(l)}}$ will be denoted by $\bm{u}^{(k)}$ and $\bm{J}^{(kl)}$, respectively, and we will sometimes express, for example, $\bm{u}$ as $(\bm{u}^{(1)},\bm{u}^{(2)})$.

Now, we can describe the oracle property.
The original definition of the oracle property is introduced by \cite{FanLi01} and is defined as follows:
\begin{description}
\item[(sparsity):] ${\rm P}(\hat{\bm{\beta}}_{\lambda}^{(1)}=\bm{0})\to1$ and
\item[(asymptotic normality):] $\sqrt{n}(\hat{\bm{\beta}}_{\lambda}^{(2)}-\bm{\beta}^{*(2)})\stackrel{{\rm d}}{\to}{\rm N}(\bm{0},\bm{J}^{(22)})$.
\end{description}
the estimators of the zero coefficients are exactly $\bm{0}$ with probability converging to 1 and the estimators of the non-zero coefficients have the same asymptotic distribution that they would have if the zero coefficients were known.
Instead of the sparsity, a variable selection consistency, i.e., ${\rm P}(\hat{\bm{\beta}}_{\lambda}^{(2)}\neq \bm{0})\to1$, is often mentioned in the oracle property.
In the following, we show the above three properties for our model.

\subsection{Sparsity}
\label{sec_sparsity}
To establish the oracle property of the estimator in (\ref{eq;est_n}) with $\gamma_{0}\in(1,2)$, we first show the sparsity.
Let us define a random function as follows:
\begin{align*}
\mathbb{G}_{n}(\bm{u})
=n^{-1}\sum_{i=1}^{n}\{g_{i}(\bm{\beta}^{*})-g_{i}(\bm{\beta}^{*}+\bm{u})\}+\sum_{j=1}^{p}\{\eta_{\lambda_{n}}(\beta_{j}^{*}+u_{j})-\eta_{\lambda_{n}}(\beta^{*}_{j})\}
\end{align*}
Note that the minimizer of $\mathbb{G}_{n}(\bm{u})$ is given by $\hat{\bm{u}}=\hat{\bm{\beta}}_{\lambda}-\bm{\beta}^{*}$.
By the same argument as in \citet{Rad05} or \citet{UmeNin15}, we see that
\begin{align*}
0
\geq\mathbb{G}_{n}(\hat{\bm{u}})-\mathbb{G}_{n}(\bm{0})
\geq n^{-1/2}\bm{s}_{n}^{{\rm T}}\hat{\bm{u}}+\hat{\bm{u}}^{{\rm T}}\bm{J}_{n}(\bm{\beta}^{\dagger})\hat{\bm{u}}/2+\sum_{j\in{\cal J}^{(2)}}\eta'_{\lambda_{n}}(\beta_{j}^{*})\hat{u}_{j}\{1+{\rm o}_{{\rm p}}(1)\},
\end{align*}
where $\bm{\beta}^{\dagger}$ is a segment from $\hat{\bm{\beta}}_{\lambda}$ to $\bm{\beta}^{*}$.
From (R3), the first term on the right-hand side in this inequality reduces to ${\rm O}_{{\rm p}}(n^{-1/2}\|\hat{\bm{u}}\|)$.
Moreover, the third term on the right-hand side in this inequality equals to 0 for sufficiently large $n$ from (P4), and thus we have $\hat{\bm{u}}={\rm O}_{{\rm p}}(n^{-1/2})$ because of the positive definiteness of $\bm{J}(\bm{\beta}^{\dagger})$ for sufficiently large $n$.

Furthermore, by expressing $\mathbb{G}_{n}(\bm{u})$ by $\mathbb{G}_{n}(\bm{u}^{(1)},\bm{u}^{(2)})$, we have
\begin{align*}
0&\geq \mathbb{G}_{n}(\hat{\bm{u}}^{(1)},\hat{\bm{u}}^{(2)})-\mathbb{G}_{n}(\bm{0},\hat{\bm{u}}^{(2)}) \\
&=-n^{-1/2}\bm{s}_{n}^{(1){\rm T}}\hat{\bm{u}}^{(1)}
+\hat{\bm{u}}^{(1){\rm T}}\bm{J}^{(11)}_{n}(\bm{\beta}^{\ddagger})\hat{\bm{u}}^{(1)}/2
+\hat{\bm{u}}^{(1){\rm T}}\bm{J}^{(11)}_{n}(\bm{\beta}^{\ddagger})\hat{\bm{u}}^{(2)}
+\sum_{j\in{\cal J}^{(1)}}\eta_{\lambda_{n}}(\hat{u}_{j})
\end{align*}
where $\bm{\beta}^{\ddagger}$ is a segment from $\hat{\bm{\beta}}_{\lambda}$ to $\bm{\beta}^{*}$.
By using $\hat{\bm{u}}={\rm O}_{{\rm p}}(n^{-1/2})$ and (P2), we see that the third and fourth term on the right-hand side in the above inequality reduces to ${\rm O}_{{\rm p}}(n^{-1}\|\hat{\bm{u}}\|)$ and $\lambda_{n}\|\hat{\bm{u}}^{(1)}\|_{1}\{1+{\rm o}_{{\rm p}}(1)\}$, respectively.
Accordingly, we have
\begin{align*}
\|\hat{\bm{u}}^{(1)}\|^{2}+\lambda_{n}\|\hat{\bm{u}}^{(1)}\|_{1}\{1+{\rm o}_{{\rm p}}(1)\}\leq {\rm O}_{{\rm p}}(n^{-1/2}\|\hat{\bm{u}}^{(1)}\|).
\end{align*}
Then, this implies that
\begin{align*}
n^{(\gamma_{0}-1)/2}\lambda\|n^{1/2}\hat{\bm{u}}^{(1)}\|_{1}\leq {\rm O}_{{\rm p}}(\|n^{1/2}\hat{\bm{u}}^{(1)}\|),
\end{align*}
and thus the sparsity of the $\ell_{1}$-type regularized estimator with $1<\gamma_{0}<2$ follows from $n^{1/2}\hat{\bm{u}}^{(1)}={\rm O}_{{\rm p}}(1)$ and $n^{(\gamma_{0}-1)/2}\to\infty$.
Note that the sparsity is not ensured when $\gamma_{0}=1$ because the above inequality does not contradict.

\begin{thm}[Sparsity]\label{thm;sparsity}
Let $\gamma_{0}\in(1,2)$.
Under the conditions (C1), (C2) and (P1)--(P4), the $\ell_{1}$-type regularized estimator has the sparsity.
\end{thm}

\subsection{Asymptotic distribution}
\label{sec_distribution}

By using Theorem \ref{thm;sparsity}, we can derive an asymptotic distribution of the estimator defined in (\ref{eq;est_n}).
In the case of $\gamma_{0}=1$, almost the same result as in \citet{UmeNin15} is appeared as shown bellow.
To establish the asymptotic distribution of the estimator, we use the same argument as in \citet{Zou06} except for the case of $\gamma_{0}=1$.

First, we consider the case of $\gamma_{0}\in(1,2)$.
Let us denote the objective function in (\ref{eq;est_n}) by $\mathbb{H}_{n}(\bm{\beta})$, that is,
\begin{align*}
\mathbb{H}_{n}(\bm{\beta})
=-\sum_{i=1}^{n}g_{i}(\bm{\beta})+n\sum_{j=1}^{n}\eta_{\lambda_{n}}(\beta_{j}).
\end{align*}
From Lemma \ref{lem;consistency} and Theorem \ref{thm;sparsity}, we see that $\hat{\bm{\beta}}_{\lambda}$ is $n^{1/2}$-consistent and that $\hat{\bm{\beta}}_{\lambda}^{(2)}$ stays away from 0 for sufficiently large $n$.
Thus, it satisfies the likelihood equation:
\begin{align}
\left.\frac{\partial\mathbb{H}_{n}(\bm{\beta})}{\partial\bm{\beta}^{(2)}}\right|_{\bm{\beta}=\hat{\bm{\beta}}_{\lambda}}
=-\sum_{i=1}^{n}g'^{(2)}_{i}(\hat{\bm{\beta}}_{\lambda})+n\bm{\eta}'_{n}(\hat{\bm{\beta}}_{\lambda}^{(2)})=\bm{0}
\label{eq;likelihood_eq}
\end{align}
with probability converging to 1, where $\bm{\eta}'_{n}(\hat{\bm{\beta}}_{\lambda}^{(2)})=(\eta'_{\lambda_{n}}(\hat{\beta}_{\lambda,j}))_{j\in{\cal J}^{(2)}}$.
Using the Taylor's theorem, we have
\begin{align*}
\bm{\eta}'_{n}(\hat{\bm{\beta}}_{\lambda}^{(2)})
=\bm{\eta}'_{n}(\bm{\beta}^{*(2)})+H_{n}(\hat{\bm{\beta}}_{\lambda}^{(2)}-\bm{\beta}^{*(2)}),
\end{align*}
where $H_{n}$ is a diagonal matrix whose $j$-th diagonal element is $\eta''_{\lambda_{n}}(\tilde{\beta}_{j})$ and $\tilde{\beta}_{j}$ is a segment from $\hat{\beta}_{\lambda,j}$ to $\beta^{*}_{j}$ for $j\in{\cal J}^{(2)}$.
Because of the $n^{1/2}$-consistency of $\hat{\bm{\beta}}_{\lambda}$, we see that $\bm{\eta}'^{(2)}_{n}(\hat{\bm{\beta}}_{\lambda}^{(2)})={\rm o}_{{\rm p}}(n^{-1/2})$ for sufficiently large $n$ from (P4) and (P5).
Moreover, $g'^{(2)}_{i}(\hat{\bm{\beta}}_{\lambda})$ can be expressed as
\begin{align*}
g'^{(2)}_{i}(\hat{\bm{\beta}}_{\lambda})
=g'^{(2)}(\bm{\beta}^{*})+g''^{(21)}_{i}(\bm{\beta}^{*})\hat{\bm{\beta}}_{\lambda}^{(1)}+g''^{(22)}_{i}(\bm{\beta}^{*})(\hat{\bm{\beta}}_{\lambda}^{(2)}-\bm{\beta}^{*(2)})+{\rm o}_{{\rm p}}(1),
\end{align*}
by the Taylor expansion.
From Theorem \ref{thm;sparsity} the second term on the left-hand side of the above equality equals to $\bm{0}$ with probability converging to 1, and thus (\ref{eq;likelihood_eq}) reduces to
\begin{align*}
-n^{1/2}\bm{s}_{n}^{(2)}+n\bm{J}_{n}^{(22)}(\hat{\bm{\beta}}_{\lambda}^{(2)}-\bm{\beta}^{*(2)})+{\rm o}_{{\rm p}}(1)+{\rm o}_{{\rm p}}(n^{1/2})=\bm{0}.
\end{align*}
Now we obtain the following theorem:

\begin{thm}[Asymptotic distribution]\label{adist_l1}
Let $\bm{J}^{(1|2)}=\bm{J}^{(11)}-\bm{J}^{(12)}\bm{J}^{(22)-1}\bm{J}^{(21)}$, $\bm{s}_{n}^{(1|2)}=\bm{s}_{n}^{(1)}-\bm{J}^{(12)}\bm{J}^{(22)-1}\bm{s}_{n}^{(2)}$ and 
\begin{align}
\hat{\bm{u}}_{n}^{(1)}
=\underset{\bm{u}^{(1)}}{{\rm argmin}}\{\bm{u}^{(1){\rm T}}\bm{J}^{(1|2)}\bm{u}^{(1)}/2-\bm{u}^{(1){\rm T}}\bm{s}_{n}^{(1|2)}+\lambda\|\bm{u}^{(1)}\|_{1}\}.
\label{eq;u1_ext}
\end{align}
Under the conditions (C1), (C2), and (P1)--(P5), we have
\begin{align*}
n^{1/2}(\hat{\bm{\beta}}_{\lambda}^{(2)}-\bm{\beta}^{*(2)})=\bm{J}^{(22)-1}\bm{s}_{n}^{(2)}+{\rm o}_{{\rm p}}(1)
\end{align*}
when $\gamma_{0}\in(1,2)$.
Moreover, we have
\begin{align}
n^{1/2}\hat{\bm{\beta}}_{\lambda}^{(1)}=\hat{\bm{u}}_{n}^{(1)}+{\rm o}_{{\rm p}}(1)
\label{eq;asy_l11}
\end{align}
and
\begin{align}
n^{1/2}(\hat{\bm{\beta}}_{\lambda}^{(2)}-\bm{\beta}^{*(2)})
=\bm{J}^{(22)-1}(\bm{s}_{n}^{(2)}-\bm{J}^{(21)}\hat{\bm{u}}_{n}^{(1)})+{\rm o}_{{\rm p}}(1)
\label{eq;asy_l12}
\end{align}
when $\gamma_{0}=1$.
\end{thm}
\noindent
Combining Theorem \ref{thm;sparsity} and \ref{adist_l1}, we see that the $\ell_{1}$-type regularized estimator has the oracle property when $\gamma_{0}\in(1,2)$.

Unlike the case of $\gamma_{0}\in(1,2)$, we can not show the oracle property of the estimator although the another asymptotic distribution can be derived.
We need using the same argument as in \citet{UmeNin15} to derive (\ref{eq;asy_l11}) and (\ref{eq;asy_l12}).
This is because that we can not show the sparsity when $\gamma_{0}=1$ and that the penalty term is not differentiable at the origin.
Note that by the convexity lemma in \citet{HjoPol93} and (R3), we see that $\hat{\bm{u}}^{(1)}$ converges in distribution to
\begin{align}
\hat{\bm{u}}^{(1)}
=\underset{\bm{u}^{(1)}}{{\rm argmin}}\{\bm{u}^{(1){\rm T}}\bm{J}^{(1|2)}\bm{u}^{(1)}/2-\bm{u}^{(1){\rm T}}\bm{s}^{(1|2)}+\lambda\|\bm{u}^{(1)}\|_{1}\}.
\label{eq;u_hat_ext}
\end{align}
The proof of (\ref{eq;asy_l11}) and (\ref{eq;asy_l12}) are given in Appendix \ref{app_pr1}.

\subsection{Variable selection consistency}
\label{sec_vsc}
The variable selection consistency, ${\rm P}(\hat{\bm{\beta}}_{\lambda}^{(2)}\neq \bm{0})\to1$, is equivalent to ${\rm P}(\hat{\cal J}^{(2)}={\cal J}^{(2)})\to1$, where $\hat{\cal J}^{(2)}=\{j;\;\hat{\beta}_{\lambda,j}\neq 0\}$ is a so-called active set.
From Lemma \ref{lem;consistency}, we see that for any $j\in{\cal J}^{(2)}$, ${\rm P}(j\in\hat{\cal J}^{(2)})\to 1$ and thus ${\rm P}(\hat{\cal J}^{(2)}\supset{\cal J}^{(2)})\to1$. 
Therefore, to establish the variable selection consistency, it suffices to show that
\begin{align}
{\rm P}(j\in\hat{\cal J}^{(2)})\to 0\;\;\;{\rm for\;any}\;\;\;j\not\in {\cal J}^{(2)}.
\label{eq;subset}
\end{align}

To show this, let us consider the event $j\in\hat{\cal J}^{(2)}$.
As mentioned in Section \ref{sec_distribution}, $\hat{\beta}_{\lambda,j}$ satisfies the likelihood equation with probability converging to 1, that is,
\begin{align*}
-\sum_{i=1}^{n}\left.\frac{\partial g_{i}(\bm{\beta})}{\partial \beta_{j}}\right|_{\bm{\beta}=\hat{\bm{\beta}}_{\lambda}}+n\eta'_{\lambda_{n}}(\hat{\beta}_{\lambda,j})
=0
\end{align*}
for any $j\in\hat{\cal J}^{(2)}$.
By using the Taylor's theorem, we see that
\begin{align*}
-\sum_{i=1}^{n}\left.\frac{\partial g_{i}(\bm{\beta})}{\partial \beta_{j}}\right|_{\bm{\beta}=\hat{\bm{\beta}}_{\lambda}}
=-n^{1/2}s_{n,j}+n\sum_{k=1}^{p}\bm{J}_{n}(\bm{\beta}^{\dagger})_{jk}(\hat{\beta}_{\lambda,k}-\beta_{k}^{*}),
\end{align*}
and then we have
\begin{align}
s_{n,j}-\sum_{k=1}^{p}\bm{J}_{n}(\bm{\beta}^{\dagger})_{jk}\{n^{1/2}(\hat{\beta}_{\lambda,k}-\beta_{k}^{*})\}
=n^{1/2}\eta'_{\lambda_{n}}(\hat{\beta}_{\lambda,j}),
\label{eq;sign}
\end{align}
where $\bm{\beta}^{\dagger}$ is a segment from $\hat{\bm{\beta}}_{\lambda}$ to $\hat{\bm{\beta}}^{*}$.
From (R2), $\bm{s}_{n}={\rm O}_{{\rm p}}(1)$ and $n^{1/2}$-consistency of $\hat{\bm{\beta}}_{\lambda}$, the left-hand side of (\ref{eq;sign}) reduces to ${\rm O}_{{\rm p}}(1)$.
Moreover, the right-hand side of (\ref{eq;sign}) goes to infinity since $j\not\in{\cal J}^{(2)}$, (\ref{eq;dif}) and $n^{1/2}\lambda_{n}\to\infty$.
Therefore, we have that ${\rm P}(j\in\hat{\cal J}^{(2)})$ is bounded above by
\begin{align*}
{\rm P}\left(s_{n,j}-\sum_{k=1}^{p}\bm{J}_{n}(\bm{\beta}^{\dagger})_{jk}\{n^{1/2}(\hat{\beta}_{\lambda,k}-\beta_{k}^{*})\}
=n^{1/2}\eta'_{\lambda_{n}}(\hat{\beta}_{\lambda,j})\right)
\end{align*}
and this converges to 0 for any $j\not\in{\cal J}^{(2)}$.
As a result, we obtain the variable selection consistency of the $\ell_{1}$-type reguralized estimator.
\begin{thm}[variable selection consistency]
\label{thm;vsc}
Under the conditions (C1), (C2) and (P1)--(P5), the $\ell_{1}$-type regularized estimator has the variable selection consistency when $\gamma_{0}\in(1,2)$.
\end{thm}

\section{Information criterion}
\label{sec_ic}
From the perspective of prediction, model selection using the AIC aims to minimize twice the Kullback-Leibler divergence (\citealt{KulLei51}) between the true distribution and the estimated distribution, 
\begin{align*}
2\tilde{{\rm E}}\left[\sum_{i=1}^{n}\tilde{g}_{i}(\bm{\beta}^{*})\right]-2\tilde{{\rm E}}\left[\sum_{i=1}^{n}\tilde{g}_{i}(\hat{\bm{\beta}}_{\lambda})\right],
\end{align*}
where $(\tilde{\bm{y}}_{1},\tilde{\bm{y}}_{2},\ldots,\tilde{\bm{y}}_{n})$ is a copy of $(\bm{y}_{1},\bm{y}_{2},\ldots,\bm{y}_{n})$; in other words, $(\tilde{\bm{y}}_{1},\tilde{\bm{y}}_{2},\ldots,\tilde{\bm{y}}_{n})$ has the same distribution as $(\bm{y}_{1},\bm{y}_{2},\ldots,\bm{y}_{n})$ and is independent of $(\bm{y}_{1},\bm{y}_{2},\ldots,\bm{y}_{n})$.
In addition, $\tilde{g}_{i}(\bm{\beta})$ and $\tilde{{\rm E}}$ denote a log-likelihood function based on $\tilde{\bm{y}}_{i}$, that is, $\log f(\tilde{\bm{y}}_{i};\bm{X}_{i}\bm{\beta})$, and the expectation with respect to only $(\tilde{\bm{y}}_{1},\tilde{\bm{y}}_{2},\ldots,\tilde{\bm{y}}_{n})$, respectively.
Because the first term is a constant, i.e., it does not depend on the model selection, we only need to consider the second term, and then the AIC is defined as an asymptotically biased estimator for it (\citealt{aka73}).
A simple estimator of the second term in our setting is $-2\sum_{i=1}^{n}g_{i}(\hat{\bm{\beta}}_{\lambda})$, but it underestimates the second term.
Consequently, we will minimize the bias correction, 
\begin{align}
-2\sum_{i=1}^{n}g_{i}(\hat{\bm{\beta}}_{\lambda})
+2{\rm E}\left[\sum_{i=1}^{n}g_{i}(\hat{\bm{\beta}}_{\lambda})-\tilde{{\rm E}}\left[\sum_{i=1}^{n}\tilde{g}_{i}(\hat{\bm{\beta}}_{\lambda})\right]\right],
\label{eq;ic}
\end{align}
in AIC-type information criteria (see \citealt{KonKit08}).
Because the expectation in (\ref{eq;ic}), i.e., the bias term, depends on the true distribution, it cannot be explicitly given in general; thus, we will evaluate it asymptotically in the same way as was done for the AIC.

For the Lasso, \citet{EfrHasJohTib04} and \citet{ZouHasTib07} developed the $C_{p}$-type information criterion as an unbiased estimator of the prediction squared error in a Gaussian linear regression setting, in other words, a finite correction of the AIC (\citealt{Sug78}) in a Gaussian linear setting with a known variance.
Unfortunately, since Stein's unbiased risk estimation theory (\citealt{Ste81}) was used for deriving this criterion, it was difficult to extend this result to other models.
In that situation, \citet{NinKaw14} relied on statistical asymptotic theory and extended the result to generalized linear models based on the asymptotic distribution of the Lasso estimator.
Because the Lasso shrinks the estimator to the zero vector too strongly, the efficiency of the parameter estimation is not necessarily large, and then \citet{UmeNin15} extended the result to non-concave penalized likelihood methods such as the SCAD and the MCP.
However they did not consider the consistency of the bias correction term of the AIC, and in fact, it is difficult to derive a consistent estimator of it in their setting especially in case of using $\ell_{1}$-type penalty.
In this study, we derive an AIC-type information criterion based on the same argument as in \citet{NinKaw14} and \citet{UmeNin15}, and construct a consistent estimator of the bias correction term of the AIC by using Theorem \ref{adist_l1} and \ref{thm;vsc}.

The bias term in (\ref{eq;ic}) can be rewritten as the expectation of
\begin{align}
\sum_{i=1}^{n}\{g_{i}(\hat{\bm{\beta}}_{\lambda})-g_{i}(\bm{\beta}^{*})\}
-\sum_{i=1}^{n}\{\tilde{g}_{i}(\hat{\bm{\beta}}_{\lambda})-\tilde{g}_{i}(\bm{\beta}^{*})\},
\label{eq;bias}
\end{align}
so we can derive an AIC by evaluating ${\rm E}[z^{{\rm limit}}]$, where $z^{{\rm limit}}$ is the limit to which (\ref{eq;bias}) converges in distribution. We call ${\rm E}[z^{{\rm limit}}]$ an asymptotic bias.
Here, we will develop an argument by setting $\gamma_{0}\in(1,2)$.

Using the Taylor's theorem, the first term in (\ref{eq;bias}) can be expressed as
\begin{align*}
n^{1/2}(\hat{\bm{\beta}}_{\lambda}-\bm{\beta}^{*})^{{\rm T}}\bm{s}_{n}
-n(\hat{\bm{\beta}}_{\lambda}-\bm{\beta}^{*})^{{\rm T}}\bm{J}_{n}(\bm{\beta}^{\dagger})(\hat{\bm{\beta}}_{\lambda}-\bm{\beta}^{*})/2,
\end{align*}
where $\bm{\beta}^{\dagger}$ is a vector on the segment from $\hat{\bm{\beta}}_{\lambda}$ to $\bm{\beta}^{*}$.
From Theorem \ref{thm;sparsity}, \ref{adist_l1} and (R2), we see that it can be rewritten by
\begin{align*}
\bm{s}_{n}^{(2){\rm T}}\bm{J}^{(22)-1}\bm{s}_{n}^{(2)}-\bm{s}_{n}^{(2){\rm T}}\bm{J}^{(22)-1}\bm{s}_{n}^{(2)}/2+{\rm o}_{{\rm p}}(1),
\end{align*}
and this converges in distribution to
\begin{align*}
\bm{s}^{(2){\rm T}}\bm{J}^{(22)-1}\bm{s}^{(2)}-\bm{s}^{(2){\rm T}}\bm{J}^{(22)-1}\bm{s}^{(2)}/2.
\end{align*}
On the other hand, the second term in (\ref{eq;bias}) can be expressed as
\begin{align*}
n^{1/2}(\hat{\bm{\beta}}_{\lambda}-\bm{\beta}^{*})^{{\rm T}}\tilde{\bm{s}}_{n}
-n(\hat{\bm{\beta}}_{\lambda}-\bm{\beta}^{*})^{{\rm T}}\bm{J}_{n}(\bm{\beta}^{\ddagger})(\hat{\bm{\beta}}_{\lambda}-\bm{\beta}^{*})/2,
\end{align*}
where $\bm{\beta}^{\ddagger}$ is a vector on the segment from $\hat{\bm{\beta}}_{\lambda}$ to $\bm{\beta}^{*}$.
By the same way as in the above, it can be rewritten by
\begin{align*}
\bm{s}_{n}^{(2){\rm T}}\bm{J}^{(22)-1}\tilde{\bm{s}}_{n}^{(2)}-\bm{s}_{n}^{(2){\rm T}}\bm{J}^{(22)-1}\bm{s}_{n}^{(2)}/2+{\rm o}_{{\rm p}}(1),
\end{align*}
and this converges in distribution to
\begin{align*}
\bm{s}^{(2){\rm T}}\bm{J}^{(22)-1}\tilde{\bm{s}}^{(2)}-\bm{s}^{(2){\rm T}}\bm{J}^{(22)-1}\bm{s}^{(2)}/2,
\end{align*}
where $\tilde{\bm{s}}_{n}^{(2)}$ and $\tilde{\bm{s}}^{(2)}$ is a copy of $\bm{s}_{n}^{(2)}$ and $\bm{s}^{(2)}$, respectively.
Therefore, we have
\begin{align*}
z^{{\rm limit}}
=\bm{s}^{(2){\rm T}}\bm{J}^{(22)-1}\bm{s}^{(2)}-\bm{s}^{(2){\rm T}}\bm{J}^{(22)-1}\tilde{\bm{s}}^{(2)}.
\end{align*}
Because $\bm{s}^{(2)}$ and $\tilde{\bm{s}}^{(2)}$ are independently distributed according to ${\rm N}(\bm{0},\bm{J}^{(22)})$, the asymptotic bias reduces to
\begin{align*}
{\rm E}[z^{{\rm limit}}]
={\rm E}[\bm{s}^{(2){\rm T}}\bm{J}^{(22)-1}\bm{s}^{(2)}]
=|{\cal J}^{(2)}|.
\end{align*}
Now we have the following theorem.

\begin{thm}\label{thm;abias_ext}
Under the same conditions as in Theorem \ref{adist_l1}, the asymptotic bias of the Kullback-Leibler divergence reduces to
\begin{align*}
{\rm E}[z^{{\rm limit}}]=|{\cal J}^{(2)}|
\end{align*}
when $\gamma_{0}\in(1,2)$, and 
\begin{align*}
{\rm E}[z^{{\rm limit}}]=|{\cal J}^{(2)}|+K
\end{align*}
when $\gamma_{0}=1$, where $K={\rm E}[\hat{\bm{u}}^{(1){\rm T}}\bm{s}^{(1|2)}]$ and $\hat{\bm{u}}^{(1)}$ is defined in (\ref{eq;u_hat_ext}).
\end{thm}
\noindent
In a similar way, we can prove the result when $\gamma_{0}=1$, so we omit the proof. 
The difference between the result of $\gamma_{0}\in(1,2)$ and $\gamma_{0}=1$ is occurred from the estimator has the sparsity or not.
Moreover, when $\gamma_{0}=1$ the resulting asymptotic bias is almost the same as in \citet{UmeNin15}, that is, we need to evaluate the expectation in $K$.

Because the asymptotic bias derived in Theorem \ref{thm;abias_ext} depends on an unknown value $\bm{\beta}^{*}$.
Replacing ${\cal J}^{(2)}$ by the active set $\hat{{\cal J}}^{(2)}=\{j;\;\hat{\beta}_{\lambda,j}\neq 0\}$ and $K$ by its empirical mean $\hat{K}$ obtained by generating samples from ${\rm N}(\bm{0},\bm{J}_{n}(\hat{\bm{\beta}}_{\lambda}))$, we can define the following index as an AIC for $\ell_{1}$-type regularization method:
\begin{align}
{\rm AIC}_{\lambda}^{\ell_{1}\mathchar`-{\rm type}}
 =\left\{ \begin{array}{lc}
 \displaystyle{-2\sum_{i=1}^{n}g_{i}(\hat{\bm{\beta}}_{\lambda})+2| \hat{{\cal J}}^{(2)}|},&\gamma_{0}\in(1,2) \\
 \displaystyle{-2\sum_{i=1}^{n}g_{i}(\hat{\bm{\beta}}_{\lambda})+2| \hat{{\cal J}}^{(2)}|+\hat{K}},&\gamma_{0}=1
\end{array} \right..
\label{eq;aic_l1}
\end{align}
It is immediately see that from Theorem \ref{thm;vsc} and
\begin{align*}
{\rm P}(|\hat{{\cal J}}^{(2)}|=|{\cal J}^{(2)}|)\geq {\rm P}(\hat{{\cal J}}^{(2)}={\cal J}^{(2)}),
\end{align*}
$|\hat{{\cal J}}^{(2)}|$ is a consistent estimator of the asymptotic bias when $\gamma_{0}\in(1,2)$.

\section{AIC for the Bridge estimator}
\label{sec_aic_bridge}
We can derive the same result as in Section \ref{sec_Asymptotics} for the Bridge estimator defined as
\begin{align*}
\hat{\bm{\beta}}_{\lambda}=
\underset{\bm{\beta}\in{\cal B}}{{\rm argmin}}\left\{-\sum_{i=1}^{n}g_{i}(\bm{\beta})+n\lambda_{n}\sum_{j=1}^{p}|\beta_{j}|^{\gamma}\right\},
\end{align*}
where $\gamma\in(0,1)$ is a known constant.
Let us consider the tuning parameter $\lambda_{n}=n^{(\gamma_{0}-2)/2}\lambda$ with $\gamma_{0}\in(\gamma,1]$ though $\gamma_{0}\in[1,2)$ for an $\ell_{1}$-type regularization method.
The choice of $\gamma_{0}$ is caused by the property of the Bridge penalty at the origin; that is, the derivative at the origin goes to infinity at the origin. 
This means that the Bridge penalty shrinks the estimator larger than that of $\ell_{1}$-type penalty around at the origin.

By the same way as in Section \ref{sec_sparsity} and \ref{sec_distribution}, we see that the Bridge estimator is consistent estimator of $\bm{\beta}^{*}$ and has the sparsity not only $\gamma_{0}\in(\gamma,1)$ but also $\gamma_{0}=1$ (see, \citealt{KniFu00,UmeNin15} for more details).
The asymptotic distribution of $\hat{\bm{\beta}}_{\lambda}$, however, is different between $\gamma_{0}\in(\gamma,1)$ and $\gamma_{0}=1$.
For the Bridge estimator we have the following theorem:

\begin{thm}[Sparsity and asymptotic distribution]\label{adist_bridge}
Under the conditions (C1), (C2), the Bridge estimator has the sparsity, that is, ${\rm P}(\hat{\bm{\beta}}_{\lambda}^{(1)}=\bm{0})\to1$ and has the asymptotic distribution
\begin{align*}
n^{1/2}(\hat{\bm{\beta}}_{\lambda}^{(2)}-\bm{\beta}^{*(2)})=\bm{J}^{(22)-1}\bm{s}_{n}^{(2)}+{\rm o}_{{\rm p}}(1)
\end{align*}
when $\gamma_{0}\in(\gamma,1)$, and
\begin{align}
n^{1/2}(\hat{\bm{\beta}}_{\lambda}^{(2)}-\bm{\beta}^{*(2)})
=\bm{J}^{(22)-1}(\bm{s}_{n}^{(2)}-\lambda\bm{\eta}'^{(2)})+{\rm o}_{{\rm p}}(1)
\label{eq;asy_bridge}
\end{align}
when $\gamma_{0}=1$, where $\bm{\eta}'^{(2)}=(\gamma{\rm sgn}(\beta_{j}^{*})|\beta_{j}^{*}|^{\gamma-1})_{j\in{\cal J}^{(2)}}$ and ${\rm sgn}(\beta)$ is the sign function; ${\rm sgn}(\beta)=1$ if $\beta>0$, ${\rm sgn}(\beta)=-1$ if $\beta<0$, and ${\rm sgn}(\beta)=0$ if $\beta=0$.
\end{thm}
\noindent
As a result in Theorem \ref{adist_bridge}, the Bridge estimator has the oracle property if $\gamma_{0}\in(\gamma,1)$ and remain the asymptotic bias if $\gamma_{0}=1$.

To state the variable selection consistency of the Bridge estimator, note that ${\rm P}(\hat{\cal J}^{(2)}\supset{\cal J}^{(2)})$ converges to 1 and let us consider the event $j\in\hat{\cal J}^{(2)}$ for any $j\in{\cal J}^{(2)}$.
By the same argument as in Section \ref{sec_vsc}, we see that
\begin{align*}
s_{n,j}-\sum_{k=1}^{p}\bm{J}_{n}(\bm{\beta}^{\dagger})_{jk}\{n^{1/2}(\hat{\beta}_{\lambda,k}-\beta_{k}^{*})\}
=\gamma\lambda n^{(\gamma_{0}-\gamma)/2}{\rm sgn}(\hat{\beta}_{\lambda,j})|n^{1/2}\hat{\beta}_{\lambda,j}|^{\gamma-1},
\end{align*}
instead of (\ref{eq;sign}), where $\bm{\beta}^{\dagger}$ is a segment from $\hat{\bm{\beta}}_{\lambda}$ to $\hat{\bm{\beta}}^{*}$.
Because the left-hand side of this equality reduces to ${\rm O}_{{\rm p}}(1)$ from the same reason as in Section \ref{sec_vsc} and the right-hand side of this equality goes to infinity from $\gamma_{0}>\gamma$ and $\hat{\bm{\beta}}={\rm O}_{{\rm p}}(1)$, we conclude that the variable selection consistency hold for the Bridge estimator when $\gamma_{0}\in(\gamma,1]$.
\begin{thm}[variable selection consistency]
\label{thm;vsc_bridge}
Under the conditions (C1), (C2), the Bridge estimator has the variable selection consistency when $\gamma_{0}\in(\gamma,1]$.
\end{thm}
Finally, we obtain the asymptotic bias of the Kullback-Leibler divergence based on the Bridge estimator by the same way as in Section \ref{sec_ic}.
\begin{thm}\label{thm;bias_bridge}
Under the conditions (C1), (C2), the asymptotic bias of the Kullback-Leibler divergence reduces to
\begin{align*}
{\rm E}[z^{{\rm limit}}]=|{\cal J}^{(2)}|,
\end{align*}
when $\gamma_{0}\in(\gamma,1]$.
\end{thm}
We can immediately see that the active set of the Bridge estimator, i.e., $|\hat{\cal J}^{(2)}|=\{j;\;\hat{\beta}_{\lambda,j}\neq 0\}$ is a consistent estimator of the asymptotic bias in Theorem \ref{thm;bias_bridge}, and thus we can define the following index as an AIC for the Bridge estimation method:
\begin{align}
{\rm AIC}_{\lambda}^{{\rm Bridge}}
 =-2\sum_{i=1}^{n}g_{i}(\hat{\bm{\beta}}_{\lambda})+2| \hat{{\cal J}}^{(2)}|
\label{eq;aic_bridge}
\end{align}
This criterion has the same form as in \cite{UmeNin15}.


\section{Conclusion and future works}\label{sec_conclusion}
In this paper, we have derived the oracle property of the estimator defined in (\ref{eq;est_n}), and developed the AIC-type information criterion by the same way as in \citet{NinKaw14} and \citet{UmeNin15}.
The resulting AIC has had almost the same asymptotic bias for the Bridge penalty or $\ell_{1}$-type penalty with $\gamma_{0}=1$, although the AIC has became more simply for the $\ell_{1}$-type penalty with $\gamma_{0}\in(1,2)$.
This phenomenon has occurred from the fact that whether the estimator has the sparsity.
We have also discussed the consistency of the AIC in (\ref{eq;aic_l1}) and (\ref{eq;aic_bridge}), and found that the the consistent estimator of the asymptotic bias of the AIC is just the active set of the estimator when the variable selection consistency hold.
It is interesting that the asymptotic behavior of the Bridge estimator and the $\ell_{1}$-type regularized estimator are different by the value of $\gamma_{0}$.
That is, we require $\gamma<\gamma_{0}<1$ for the Bridge estimator and $1<\gamma_{0}<2$ for the $\ell_{1}$-type regularized estimator to assure the oracle property.
This is because the Bridge penalty has a property that the derivative at the origin diverges and that it is not bounded uniformly, although the $\ell_{1}$-type penalty behaves like the Lasso around the origin and is uniformly bounded.
Moreover, the only behavior around at the origin is important for the Bridge penalty while the uniform boundedness and divergence of $n\lambda_{n}$ are important for the $\ell_{1}$-type penalty to assure the sparsity.
Although we have only considered the Bridge penalty for $\ell_{\gamma}$-type regularization method in this paper, we can generalize the result for a general $\ell_{\gamma}$-type penalty to assume about the behavior of the first derivative of such penalty. 


In this study, we derived the AIC based on statistical asymptotic theory for which the dimension of the parameter vector is fixed and the sample size diverges.
On the other hand, it is becoming important to analyze high-dimensional data wherein the dimension of the parameter vector is comparable to the sample size.
Also for such high-dimensional data, we expect that the AIC-type information criterion will work well from the viewpoint of efficiency.
In fact, \citet{ZhaLiTsa10} has shown that, when the dimension of the parameter vector increases with the sample size, their criterion close to the proposed information criterion has an asymptotic loss efficiency in a sparse setting under certain conditions.
It will be important in terms of both theory and practice to show that the proposed information criterion has a similar asymptotic property.

\appendix

\section{Proof of (\ref{eq;asy_l11}) and (\ref{eq;asy_l12})}
\label{app_pr1}
To prove (\ref{eq;asy_l11}) and (\ref{eq;asy_l12}), we use the following lemma:
\begin{lem}
\label{lem;HjoPol}
Suppose that $\phi_{n}(\bm{u})$ is a strictly convex random function that is approximated by $\tilde{\phi}_{n}(\bm{u})$.
Let $\psi(\bm{u})$ be a convex continuous function such that $\psi_{n}(\bm{u})$ converges to $\psi(\bm{u}^{\dagger})$ uniformly over $\bm{u}$ in any compact set.
In addition, for
\begin{align*}
\nu_{n}(\bm{u})=\phi_{n}(\bm{u})+\psi_{n}(\bm{u})\;\;\;\;\; and \;\;\;\;\;
\tilde{\nu}_{n}(\bm{u})=\tilde{\phi}_{n}(\bm{u})+\psi(\bm{u}),
\end{align*}
let $\bm{u}_{n}$ and $\tilde{\bm{u}}_{n}$ be the minimizer of $\nu_{n}(\bm{u})$ and $\tilde{\nu}_{n}(\bm{u})$, respectively.
Then, for any $\varepsilon\;(>0)$, $\delta\;(>0)$ and $\xi\;(>\delta)$, we have
\begin{align}
{\rm P}(|\bm{u}_{n}-\tilde{\bm{u}}_{n}|\geq \delta)
\leq {\rm P}\left(2\Delta_{n}(\delta)+\varepsilon\geq \Upsilon_{n}(\delta)\right)
+{\rm P}(|\bm{u}_{n}-\tilde{\bm{u}}_{n}|\geq \xi),
\label{eq;lem3_ineq}
\end{align}
where
\begin{align}
\Delta_{n}(\delta)=\sup_{|\bm{u}-\tilde{\bm{u}}_{n}|\leq \delta}|\nu_{n}(\bm{u})-\tilde{\nu}_{n}(\bm{u})|\;\;\;\;\; and \;\;\;\;\;
\Upsilon_{n}(\delta)=\inf_{|\bm{u}-\tilde{\bm{u}}_{n}|=\delta}\tilde{\nu}_{n}(\bm{u})-\tilde{\nu}_{n}(\tilde{\bm{u}}_{n}).
\label{eq;lem3_check}
\end{align}
\end{lem}
\noindent
Lemma \ref{lem;HjoPol} is almost the same as in \citet{HjoPol93} or \citet{UmeNin15}.

Now we can prove (\ref{eq;asy_l11}) and (\ref{eq;asy_l12}).
Let us define the random function as follows:
\begin{align*}
\nu_{n}(\bm{u})=\phi_{n}(\bm{u})+\psi_{n}(\bm{u})
\end{align*}
where
\begin{align*}
\phi_{n}(\bm{u})
=\sum_{i=1}^{n}\{g_{i}(\bm{\beta}^{*})-g_{i}(\bm{\beta}^{*}+n^{-1/2}\bm{u})\}
\end{align*}
and
\begin{align*}
\psi_{n}(\bm{u})
=n\sum_{j=1}^{p}\{\eta_{\lambda_{n}}(\beta_{j}^{*}+n^{-1/2}u_{j})-\eta_{\lambda_{n}}(\beta^{*}_{j})\}
\end{align*}
Note that the minimizer of $\nu_{n}(\bm{u})$ is given by $\bm{u}_{n}=(\bm{u}_{n}^{(1)},\bm{u}_{n}^{(2)})=(n^{1/2}\hat{\bm{\beta}}_{\lambda}^{(1)},n^{1/2}(\hat{\bm{\beta}}_{\lambda}^{(2)}-\bm{\beta}^{*(2)}))$.
By the Taylor's theorem, $\phi_{n}(\bm{u})$ can be approximated by
\begin{align*}
\tilde{\phi}_{n}(\bm{u})
=-\bm{u}^{{\rm T}}\bm{s}_{n}+\bm{u}^{{\rm T}}\bm{J}\bm{u}/2.
\end{align*}
On the other hand, from condition (P2), (P4) and $\gamma_{0}=1$, we see that $\psi_{n}(\bm{u})$ converges to $\psi(\bm{u})=\lambda\|\bm{u}^{(1)}\|_{1}$ uniformly over $\bm{u}$ in any compact set.
Then, because $\tilde{\nu}_{n}(\bm{u})=\tilde{\phi}_{n}(\bm{u})+\psi(\bm{u})$ can be rewritten as
\begin{align*}
\tilde{\nu}(\bm{u})
=&\left\{\bm{u}^{(2)}-\bm{J}^{(22)-1}(\bm{s}_{n}^{(2)}-\bm{J}^{(21)}\bm{u}^{(1)})\right\}^{{\rm T}}\bm{J}^{(22)}\left\{\bm{u}^{(2)}-\bm{J}^{(22)-1}(\bm{s}_{n}^{(2)}-\bm{J}^{(21)}\bm{u}^{(1)})\right\}/2 \\
&+\bm{u}^{(1){\rm T}}\bm{J}^{(1|2)}\bm{u}^{(1)}/2
-\bm{u}^{(1){\rm T}}\bm{s}_{n}^{(1|2)}
+\lambda\|\bm{u}^{(1)}\|_{1}
-\bm{s}_{n}^{(2){\rm T}}\bm{J}^{(22)-1}\bm{s}_{n}^{(2)}/2,
\end{align*}
the minimizer of $\tilde{\nu}_{n}(\bm{u})$ is given by
$\tilde{\bm{u}}_{n}
=(\tilde{\bm{u}}_{n}^{(1)},\tilde{\bm{u}}_{n}^{(2)})
=(\hat{\bm{u}}_{n}^{(1)},\bm{J}^{(22)-1}(\bm{s}_{n}^{(2)}-\bm{J}^{(21)}\hat{\bm{u}}_{n}^{(1)}))$
where $\hat{\bm{u}}_{n}^{(1)}$ is defined by (\ref{eq;u1_ext}).

By the same argument as in \citet{UmeNin15}, $\Delta_{n}(\delta)$ converges to 0 in probability.
Next, by the definition of $\tilde{\bm{u}}_{n}^{(1)}$ we have
\begin{align*}
\bm{J}^{(1|2)}\tilde{\bm{u}}_{n}^{(1)}-\bm{\tau}_{\lambda}(\bm{s}_{n})+\lambda\bm{\gamma}=\bm{0},
\end{align*}
where $\bm{\gamma}$ is a sub-gradient of $\|\tilde{\bm{u}}_{n}^{(1)}\|_{1}$, that is, $\bm{\gamma}$ is a $|{\cal J}^{(1)}|$-dimensional vector such that $\gamma_{j}=1$ when $\hat{u}_{n,j}^{(1)}>0$, $\gamma_{j}=-1$ when $\hat{u}_{n,j}^{(1)}<0$, and $\gamma_{j}\in[-1,1]$ when $\hat{u}_{n,j}^{(1)}=0$. Thus, noting that $\tilde{\bm{u}}_{n}^{(1){\rm T}}\bm{\gamma}=\|\tilde{\bm{u}}_{n}^{(1)}\|_{1}$, we can write $\tilde{\nu}_{n}(\bm{u}^{(1)},\bm{u}^{(2)})-\tilde{\nu}_{n}(\tilde{\bm{u}}_{n}^{(1)},\tilde{\bm{u}}_{n}^{(2)})$ as
\begin{align}
&(\bm{u}^{(1)}-\tilde{\bm{u}}_{n}^{(1)})^{{\rm T}}\bm{J}^{(1|2)}(\bm{u}^{(1)}-\tilde{\bm{u}}_{n}^{(1)})/2
+\lambda\sum_{j\in{\cal J}^{(1)}}\left(|u_{j}|-\gamma_{j}u_{j}\right) \nonumber \\
&+\left\{\bm{u}^{(2)}-\bm{J}^{(22)-1}(\bm{s}_{n}^{(2)}-\bm{J}^{(21)}\bm{u}^{(1)})\right\}^{{\rm T}}\bm{J}^{(22)}\left\{\bm{u}^{(2)}-\bm{J}^{(22)-1}(\bm{s}_{n}^{(2)}-\bm{J}^{(21)}\bm{u}^{(1)})\right\}/2
\label{eq:Upsilon_n}
\end{align}
after a simple calculation. Let $\bm{w}_{1}$ and $\bm{w}_{2}$ be unit vectors such that $\bm{u}^{(1)}=\tilde{\bm{u}}_{n}^{(1)}+\zeta\bm{w}_{1}$ and $\bm{u}^{(2)}=\tilde{\bm{u}}_{n}^{(2)}+(\delta^{2}-\zeta^{2})^{1/2}\bm{w}_{2}$, where $0\leq \zeta\leq \delta$. Then, letting $\rho^{(22)}$ and $\rho^{(1|2)}\;(>0)$ be half the smallest eigenvalues of $\bm{J}^{(22)}$ and $\bm{J}^{(1|2)}$, respectively, it follows that
\begin{align*}
\Upsilon_{n}(\delta)
\geq \min_{0\leq \zeta\leq \delta}\left\{ \rho^{(1|2)}\zeta^{2}+\rho^{(22)}|(\delta^{2}-\zeta^{2})^{1/2}\bm{w}_{2}+\zeta\bm{J}^{(22)-1}\bm{J}^{(21)}\bm{w}_{1}|^{2}\right\}
>0
\end{align*}
because the second term in (\ref{eq:Upsilon_n}) is non-negative.
Thus the first term on the right-hand side in (\ref{eq;lem3_ineq}) converges to 0.
In addition, it is easy to see that the second term on the right-hand side in (\ref{eq;lem3_ineq}) can be made arbitrarily small by considering a sufficiently large $\xi$ because $(\bm{u}_{n}^{(1)},\bm{u}_{n}^{(2)})={\rm O}_{{\rm p}}(1)$ and $(\tilde{\bm{u}}_{n}^{(1)},\tilde{\bm{u}}_{n}^{(2)})={\rm O}_{{\rm p}}(1)$ from $\hat{\bm{\beta}}_{\lambda}={\rm O}_{{\rm p}}(n^{-1/2})$. 
As a consequence, we obtain (\ref{eq;asy_l11}) and (\ref{eq;asy_l12}).

  \bibliographystyle{jecon} 
  \bibliography{myref}





\end{document}